\documentclass{emulateapj}
\begin{document}

\shorttitle{FIREBALLS IN A COMA GALAXY}
\shortauthors{YOSHIDA ET AL.}

\title{Strange filamentary structures (``fireballs'') around a merger galaxy in 
the Coma cluster of galaxies\altaffilmark{1}}
\author{
Michitoshi Yoshida\altaffilmark{2},
Masafumi Yagi\altaffilmark{3},
Yutaka Komiyama\altaffilmark{3,4},
Hisanori Furusawa\altaffilmark{4},
Nobunari Kashikawa\altaffilmark{3},
Yusei Koyama\altaffilmark{5},
Hitomi Yamanoi\altaffilmark{3,6},
Takashi Hattori\altaffilmark{4}
and
Sadanori Okamura\altaffilmark{5,7}
}

\altaffiltext{1}{Based on data collected at the Subaru Telescope, operated by the
National Astronomical Observatory of Japan.}
          
\altaffiltext{2}{Okayama Astrophysical Observatory, National Astronomical 
Observatory,
Kamogata, Okayama 719-0232, Japan; yoshida@oao.nao.ac.jp.}

\altaffiltext{3}{Optical and Infrared Astronomy Division, National 
Astronomical Observatory,
Mitaka, Tokyo 181-8588, Japan.}

\altaffiltext{4}{Subaru Telescope, National Astronomical Observatory of 
Japan, 650 North
A'Ohoku Place, Hilo, HI 96720, USA.}

\altaffiltext{5}{Department of Astronomy, University of Tokyo,
Tokyo 113-0033, Japan.}

\altaffiltext{6}{Department of Astronomical Science, School of Physical Sciences,
The Graduate University for Advanced Studies (Sokendai),
National Astronomical Observatory of Japan,
Tokyo 181-8588, Japan.}

\altaffiltext{7}{Research Center for the Early Universe, University of 
Tokyo, Tokyo 113-0033, Japan.}


\begin{abstract}

We found an unusual complex of narrow blue filaments, bright blue knots, and H$\alpha$-emitting 
filaments and clouds, which morphologically resembled a complex of ``fireballs,'' extending up to 
80 kpc south from an E+A galaxy RB199 in the Coma cluster. The galaxy has a highly disturbed 
morphology indicative of a galaxy--galaxy merger remnant. The narrow blue filaments extend in 
straight shapes toward the south from the galaxy, and several bright blue knots are located at the 
southern ends of the filaments. The $R_{\rm C}$ band absolute magnitudes, half light radii and 
estimated masses of the bright knots are $\sim -12 - -13$ mag, $\sim 200-300$ pc and 
$\sim 10^{6-7}$ M$_{\odot}$, respectively.
Long, narrow H$\alpha$-emitting filaments are connected at the south edge of the knots. The 
average color of the fireballs is $B - R_{\rm C}\approx 0.5$, which is bluer than RB199 ($B - R = 
0.99$), suggesting that most of the stars in the fireballs were formed within several times $10^8$ yr. 
The narrow blue filaments exhibit almost no H$\alpha$ emission. Strong H$\alpha$ and UV emission 
appear in the bright knots. These characteristics indicate that star formation recently ceased in the 
blue filaments and now continues in the bright knots. The gas stripped by some mechanism from 
the disk of RB199 may be traveling in the intergalactic space, forming stars left along its trajectory. 
The most plausible fireball formation mechanism is ram pressure stripping by high-speed collision 
between the galaxy and the hot intra-cluster medium. The fireballs may be a snapshot of diffuse 
intra-cluster population formation, or halo star population formation in a cluster galaxy.
\end{abstract}




\section{Introduction}

Clusters of galaxies are ideal experimental laboratories for studying environmental effects on 
galaxy evolution (e.g. Bosselli \& Gavazzi 2006). Cluster galaxy density varies widely from 
edge to core, the 
intergalactic space is often filled with X-ray emitting hot gas, and the cluster's large mass forms a 
very deep gravitational potential around its center. These properties give us observational means to 
investigate how galaxy density, ambient gas--galaxy interaction, and strong gravitational drag force 
affect galaxy evolution.

Several lines of observational evidence suggest that drastic galaxy evolution occurred in clusters of 
galaxies from the redshift $z \sim 1$ to the present. Early type galaxies form a dominant population 
in the inner region of local rich clusters, and in particular, S0 galaxies are preferentially found in 
and around the cluster core region 
\citep{Dressler1980,Dressler1994,Dressler1997,Postman1984,Goto2003,VanDerWel2007}. This 
relationship---the morphology--density relationship---is also observed at high redshift ($z \sim 1$). 
However, this enhancement of the fraction of S0 galaxies around a cluster core is less remarkable 
in high redshift clusters than in local clusters \citep{Smith2005,Postman2005}. In addition, a 
higher blue galaxy fraction has been established for high-redshift clusters 
\citep{Butcher1978,Butcher1984}. Furthermore, a rapid rise in the luminosity function (LF) of 
member galaxies toward the faint end has been reported in several local rich clusters 
\citep{Popesso2005,Trentham2005,Popesso2006,Milne2007,Adami2007,Yamanoi2007}. However, 
the LFs of some rich clusters at $z \approx 0.3$ may exhibit no such rapid increase at those faint 
ends \citep{Harsono2007}. These observational facts suggest that in terms of morphology, color, 
and population, rapid evolution of galaxies occurred in clusters from $z \sim 1$ to the present.

Several environmental effects that explain the evolution of cluster galaxies have been proposed, 
including galaxy--galaxy interaction \citep{Farouki1981,Icke1985}, galaxy harassment 
\citep{Moore1996}, tidal interaction with cluster potential \citep{Henriksen1996}, ram pressure 
stripping \citep{Gunn1972,Vollmer2001,Kron2008}, turbulent viscous stripping 
\citep{Nulsen1982}, and galaxy starvation \citep{Larson1980,Bekki2002}. In addition, many 
observational studies have been made to identify the dominant environmental processes in clusters 
(cf., Bosselli \& Gavazzi 2006 and references therein). However, the kinds of physical processes 
that are dominant in rapid galaxy evolution in clusters are still unclear. 

Recently, \citet{Cortese2007} (C07) found two peculiar galaxies near the central regions of rich 
clusters Abell 1689 and Abell 2667 at $z\approx 0.2$. These galaxies have disturbed 
morphology and are associated with many small blue blobs extending toward one side of the 
galaxies. These blobs have absolute magnitudes of $M_B \approx -11 - -12$, corresponding to 
masses of $10^{6-8}$ M$_{\odot}$. C07 detected [OII] emission from some of the blobs and 
between the blobs and the host galaxies. The brightness and the size of the blobs are similar to 
those of dwarf elliptical galaxies or ultracompact dwarf galaxies \citep{Drinkwater2004}. C07 
suggested that the phenomena they found may be snapshots of the transformation of spiral galaxies 
to S0 galaxies, and may give insight into the origin of cluster faint-end populations.

Here we report the discovery of a strange complex of narrow blue filaments and knots extending 
toward one side of a disturbed E+A galaxy, RB199, in the Coma cluster. New deep optical 
broadband and narrowband (H$\alpha$) imaging observations of the central region of the Coma 
cluster revealed this structure. The morphology of the structure is very similar to that of the 
phenomena reported in C07, and the sizes and the luminosities of the knots in the structure are 
comparable to C07's blobs. This structure may be a nearby counterpart of C07's phenomena.

We assumed the cosmological parameters ($h_0$, $\Omega_m$, $\Omega_\lambda$) = (0.73, 0.24, 
0.72) and the distance modulus of the Coma cluster to be 35.05 \citep{Yagi2007}. The linear scale 
at the Coma cluster is 474 pc arcsec$^{-1}$ under this assumption.

\section{Observation and Data Reduction}

We observed a 34$\times$27 arcmin region near the Coma cluster center ($\alpha,\delta$)(J2000.) 
= (12$^h$59$^m$26$^s$, +27$^{\circ}$44\arcmin 16\arcsec) with Suprime-Cam 
\citep{Miyazaki2002} attached to the Subaru Telescope on April 28 and May 3, 2006 and May 13--
15, 2007 (Table 1). The main purpose of these observations was to study the faint-end population 
and star formation in the Coma cluster. We used two broadband filters, $B$, $R_{\rm C}$ and a 
narrowband filter (N-A-L671, hereafter H$\alpha$ NB). The H$\alpha$ NB filter is designed for 
observing H$\alpha$-emitting objects in the Coma cluster at $z = 0.0225$, and has a bell-shaped 
transmission with a central wavelength of 6712 \AA\ and a FWHM of 120 \AA.

We reduced the imaging data in the standard manner. Since some data have low S/N because of bad 
weather, we chose better S/N images to make combined images. Photometric calibration was 
performed using photometric standard stars observed on May 3, 2006 \citep{Yagi2007}. We 
corrected Galactic extinction using the extinction calculator provided on the NASA/IPAC 
Extragalactic Database (NED) Web site. The basic extinction data were obtained from 
\citet{Schlegel1998}. 
The limiting surface brightness of each band image was estimated from 1-$\sigma$ fluctuation
of sky background within a 2 arcsec diameter circular aperture. The limiting 
surface brightness and the PSF sizes of the final images are summarized in Table 1. We subtracted 
the $R_C$ image from the H$\alpha$ NB image to produce a continuum subtracted 
H$\alpha$ image (``pure (net) H$\alpha$ image''). We rescaled the $R_{\rm C}$ image before the 
subtraction using several field stars to equalize the mean flux of the stars in the $R_{\rm C}$ image 
to that of the H$\alpha$ image. The resultant limiting surface brightness (1-$\sigma$ fluctuation in 
a 2\arcsec\ circular aperture) of the pure H$\alpha$ image is $2.4\times 10^{-18}$ erg s$^{-1}$
cm$^{-2}$ arcsec$^{-2}$, if photon noise of object is negligible.

The lengths and distances given in the following sections are all projected. Since the inclination 
angles with respect to the sky plane of the features that we found are not known, we do not apply 
any projection correction. The magnitudes are expressed in the Vega system unless otherwise 
specified.

\section{Results}

A false color image made with the $B$ (blue), the $R_{\rm C}$ (green), and the H$\alpha$ NB 
(red) images around an amorphous galaxy RB199 (J2000.0, 12$^h$58$^m$43$^s$, 
+27$^{\circ}$45\arcmin 43\arcsec) is shown in Figure \ref{fig1}. The centering error of the three 
color images is less than 0.1 pixels, corresponding to 0\arcsec .02. The continuum light is not 
subtracted from the H$\alpha$ NB image in Figure \ref{fig1}. Contour maps of the $B$ image and 
the $R_{\rm C}$ image are shown in Figures \ref{fig2} and \ref{fig3}. Figure \ref{fig4} shows a 
grayscale image of the pure H$\alpha$ image. A remarkable complex of blue filaments and blobs 
and H$\alpha$-emitting clouds extends toward the south from RB199 (Figures \ref{fig1} - 
\ref{fig4}). The overall morphology of the complex resembles a group of fireballs (large meteors) 
shooting from the galaxy toward the south (see Figure \ref{fig1}). The complex is thus referred to 
as the ``fireballs.'' Figure \ref{fig5} is a $B - R_{\rm C}$ color map of RB199 and the bright part 
of the fireballs. We describe the morphological and color characteristics of RB199 and this strange 
newly observed feature in the following subsections.

\subsection{RB199}

RB199 is an E+A galaxy whose projected distance from the center of the Coma cluster is about 
18\arcmin\, corresponding to 0.5 Mpc at the cluster \citep{Poggianti2004}. RB199's amorphous 
morphology is clearly seen in Figure \ref{fig6}. The stellar system of RB199 is highly disturbed, 
indicating that this galaxy is a galaxy--galaxy merger remnant. A past merger event probably 
induced a starburst in the galaxy and now the star formation activity is in its decay phase, i.e., post-
starburst phase. E+A characteristics of the spectrum of the galaxy \citep{Poggianti2004} suggest 
that massive OB stars have already died, whereas A-type stars are still alive. The 
H$\alpha$ absorption dominant post-starburst region is distributed around the nucleus of RB199 as 
shown in the pure H$\alpha$ image (see Figure \ref{fig4}).

The main structure of RB199 can be divided into two parts: a central disturbed bright ``ellipse'' and 
a smooth ``disk'' extending west of the ellipse (Figure \ref{fig6}). The position angles of the ellipse 
and the disk are 70$^\circ$ and $-70^\circ$, respectively.

The $B - R$ color of RB199 within a 163\arcsec\ aperture diameter is $B - R = 
0.99$ \citep{Poggianti2004}. Our $B - R_{\rm C}$ color map (Figure \ref{fig5}) reveals a 
complex inner structure of color distribution in the galaxy. The central ellipse is bluer than the disk. 
The central blue region in the central ellipse has a thick V-shaped structure and a $B - R_{\rm 
C}$ color of $\approx 0.8-0.9$. The blue V-shaped structure is surrounded by slightly redder 
components whose $B - R_{\rm C}$ color and position angle are $\approx 1.0$ and $\approx 
70^{\circ}$, respectively. The disk has a $B - R_{\rm C}$ color of $\approx 1.2$. Color 
distribution of the disk is smooth and no clear color gradient appears.

These characteristics suggest that a past merger concentrated the gas and stars of the progenitor 
galaxies into the bottom of the gravitational potential of the system and induced a starburst. Some 
of the newly formed stars from the merging process might have mixed with the preexisting disk 
stars and formed the disk of RB199.

\subsection{The fireballs}

\subsubsection{Morphology}

\begin{figure}
\epsscale{0.8}
\plotone{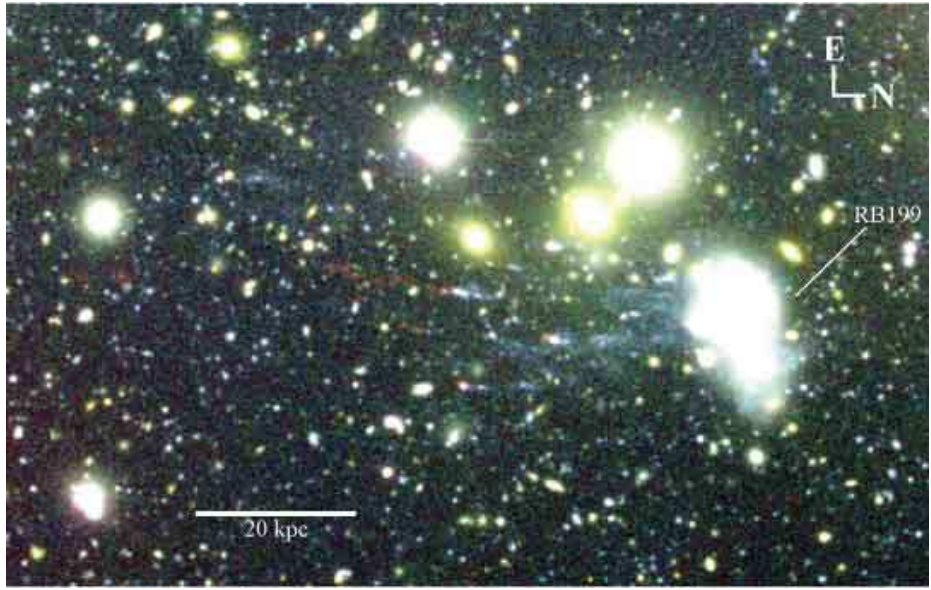}
\caption{
\label{fig1} False color ($B$ band: blue, $R_{\rm C}$ band: green, H$\alpha$ NB: red) image of area 
around RB199. North is right and east is top. RB199 is located at the right side of the image.
}
\end{figure}

\begin{figure}
\epsscale{0.6}
\plotone{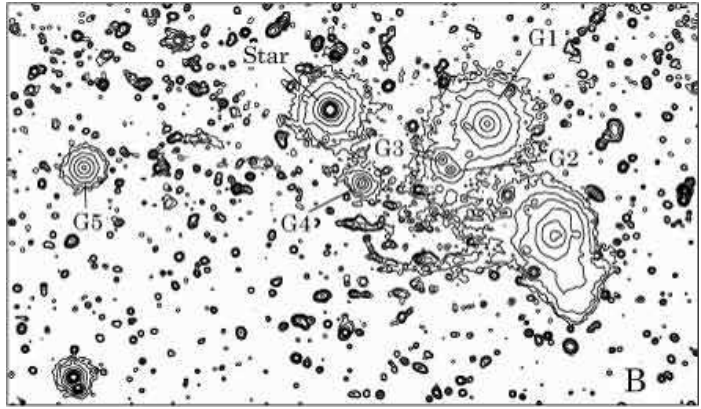}
\caption{
\label{fig2} $B$ contour map around RB199.  The orientation is the same as Figure 1. To improve 
the signal-to-noise ratio of the faint features, the original images were smoothed by $4 \times 
4$ pixels running mean to produce the contour maps in Figure \ref{fig2} and Figure \ref{fig3}. The 
lowest contour level of this image is 28.5 (AB)mag arcsec$^{-2}$ and the contour interval is 1 mag. 
``G1''--``G5'' are galaxies near RB199. The object lying to the east of the galaxy G4 is a foreground 
star.  
}
\end{figure}

\begin{figure}
\epsscale{0.6}
\plotone{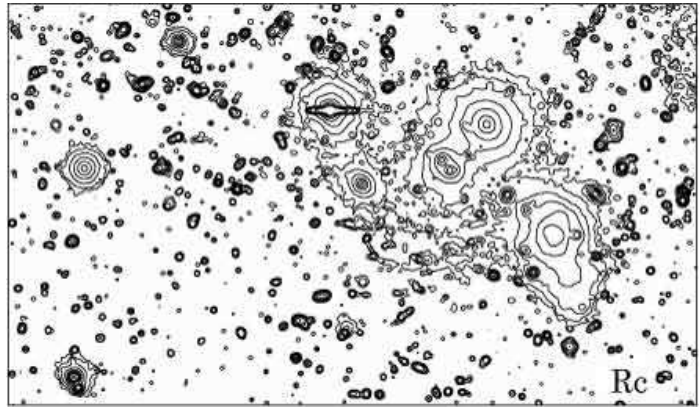}
\caption{
\label{fig3} $R_{\rm C}$ contour map around RB199. The lowest contour level is 27.8 (AB)mag 
arcsec$^{-2}$ and the contour interval is 1 mag.
}
\end{figure}

\begin{figure}
\epsscale{0.6}
\plotone{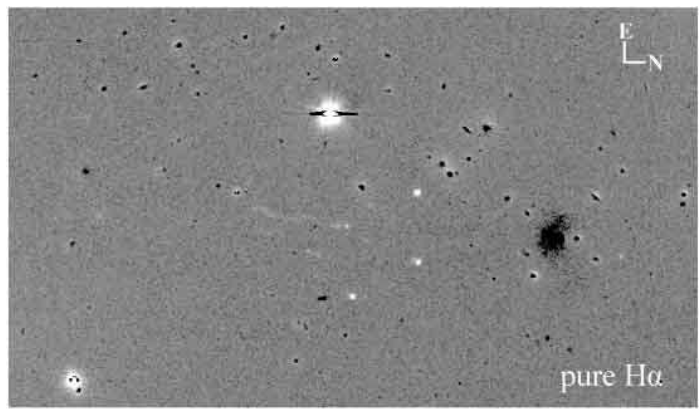}
\caption{
\label{fig4} Continuum subtracted H$\alpha$ image around RB199. Stronger H$\alpha$ is coded 
whiter. 
}
\end{figure}

\begin{figure}
\epsscale{0.6}
\plotone{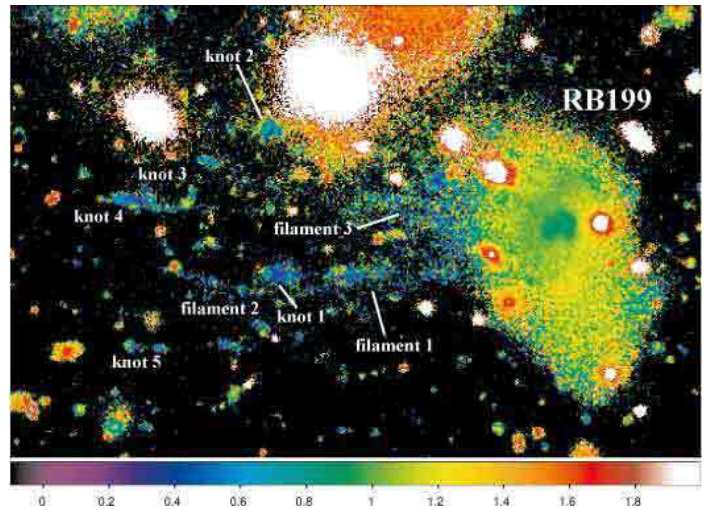}
\caption{
\label{fig5} $B - R_{\rm C}$ color map of the RB199 field. This map shows the $B - R_{\rm 
C}$ color from 0.0 (violet) to 1.8 (red) (see the color bar at the bottom of the figure).
}
\end{figure}

\begin{figure}
\epsscale{.50}
\plotone{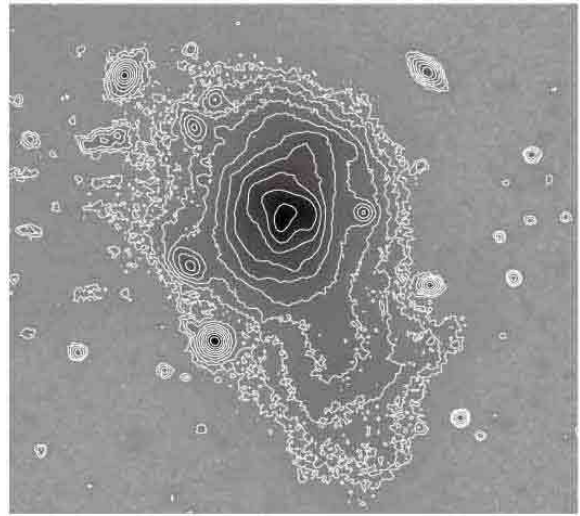}
\caption{
\label{fig6} $B$ band image of RB199 (contour map overlaid on the grayscale image). The lowest 
contour level and the contour interval are 26.5 (AB)mag arcsec$^{-2}$ and 0.5 mag, respectively.
}
\end{figure}

\begin{figure}
\epsscale{.60}
\plotone{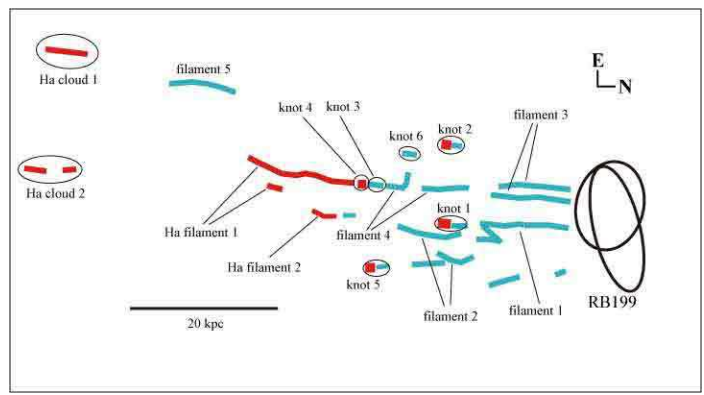}
\caption{
\label{fig7} Schematic view of the ``fireballs'' of RB199.
}
\end{figure}

\begin{figure}
\epsscale{.5}
\plotone{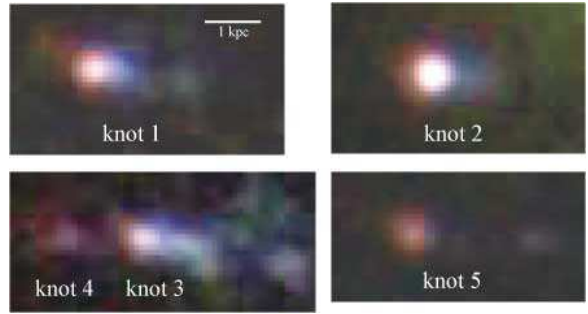}
\caption{
\label{fig8} Close-up views of the bright knots. Enlarged false-color images of the knot 1 (upper 
left), the knot 2 (upper right), the knot3 and knot 4 (lower left), and the knot 5 (lower right) are 
shown. The attribute of the color is the same as in Figure \ref{fig1}.
}
\end{figure}

The most striking feature around RB199 is a complex of strange faint blue knots and filaments and 
H$\alpha$-emitting ionized gas clouds extending toward the south from the galaxy (the ``fireballs'': 
Figures \ref{fig1} - \ref{fig4}). The whole structure of the fireballs is schematically drawn in 
Figure \ref{fig7}. We labeled several characteristic features in the fireballs, such as bright knots, 
remarkable filaments, and extended faint clouds, as ``knot 1,'' ``filament 1,'' ``cloud 1,'' and so forth 
(Figure \ref{fig7}). The enlarged false-color images of knots 1 to 5 are shown in Figure \ref{fig8}.

\begin{figure}
\epsscale{.50}
\plotone{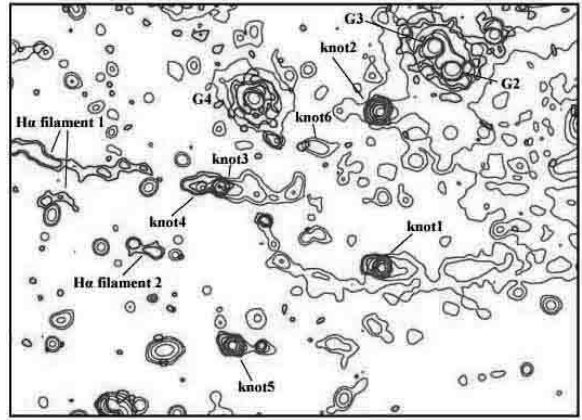}
\caption{
\label{fig9} Contour map of the pure H$\alpha$ image (red lines) is overlaid on a contour map of 
the $B$ image (blue lines). Lowest contour levels of the pure H$\alpha$ and the $B$ maps are 
$2.4\times 10^{-18}$ erg cm$^{-2}$ s$^{-1}$ arcsec$^{-2}$ and 28 (AB)mag arcsec$^{-2}$, 
respectively. Contour interval is 1 mag for both maps.
}
\end{figure}

The bright knots are located at the end of the bright blue filaments; for example, knot 1 is located at 
the southern end of filament 1. Strong H$\alpha$ emission appears at the south sides of the bright 
knots. Figures \ref{fig8} and \ref{fig9} clearly show this configuration. The intensity peaks of the 
H$\alpha$ emission are slightly displaced to the south of the $B$ band intensity peaks of the bright 
knots (see Figures \ref{fig8} and \ref{fig9}).

The blue filaments (filament 1--filament 5) consist of several filaments with narrow, long, straight, 
or slightly curved shapes, and lengths of $\sim 10 - 30$ kpc. The roots of the filaments appear to 
be connected to the central ellipse of RB199 (Figure \ref{fig1}, see also Figure \ref{fig7}). The 
most prominent filament is filament 1. It extends linearly toward knot 1, which is the brightest knot 
in the blue filaments, and connects to filament 2 at the west side of knot 1. Filament 3 contains two 
straight parallel filaments extending outward from the central ellipse. Filament 4 ends at knot 3 and 
knot 4. Knot 4 is bright in H$\alpha$ emission and is smoothly connected to the most remarkable 
H$\alpha$ feature, H$\alpha$ filament 1. H$\alpha$ filament 1 has a linear, slightly wiggled 
morphology with a length of $\approx 15$ kpc and a width of $\approx 2$ kpc.

The bright knots are spatially resolved.  Their sizes, measured at $B$ band surface brightness of
28 mag arcsec$^{-2}$, are $\sim 1 - 2$ kpc (Table 2). To estimate the half light radii (the effective 
radii, $r_{\rm e}$) of the knots, we fitted Gaussian profiles blurred by the point spread function 
(PSF) to the light profiles of the knots. Then we derived $r_{\rm e}$ using the relation $r_{\rm e} 
\approx 1.18\; \sigma$. The $r_{\rm e}$ of the knots are $\approx 200-300$ pc. Although the light 
distributions of the knots are unlikely to exhibit Gaussian profiles, the light profiles of the knots, in 
particular knots 1, 2, and 5, are well represented with Gaussians. Thus, the $r_{\rm e}$ are good 
approximations of the knot sizes.

About 50 kpc from RB199 (Figure \ref{fig1}) is a faint, diffuse filament. This filament, filament 5, 
is slightly displaced to the east from the line connecting filaments 3, 4, and H$\alpha$ filament 1 
(see also Figure \ref{fig7}). Filament 5 is marginally detected in $R_{\rm C}$ band, indicating that 
it is very blue. In addition, at the east and west of an elliptical galaxy G5 (located about 80 kpc 
from RB199) are faint, diffuse H$\alpha$ clouds (Figure \ref{fig1}). They are very faint, but are 
marginally seen in the pure H$\alpha$ image (Figure \ref{fig4}). They are also observed as faint 
red diffuse clouds in Figure \ref{fig1}.

\subsubsection{Optical photometry}

The results of the photometry of the fireballs are summarized in Table 2. The bright continuum 
knots have absolute magnitudes of $M_B\sim -12$ and $M_R\sim -12- -13$. The typical $r_{\rm 
e}$ is $\approx 200-300$ pc (see section 3.2.1). These values are similar to those of faint dwarf 
ellipticals in the Virgo cluster (e.g., Ichikawa et al. 1990) or the faint ends of local dwarf galaxies 
\citep{Belokurov2007}. Assuming a mass to luminosity ratio $M$/$L_R$ $ = $ 1 for these knots, 
their masses are $\sim 10^{6-7}$ M$_\odot$. The mean color of the blue filaments are $B - 
R_{\rm C} \approx 0.5$, corresponding to early F stars in the main sequence. The 
H$\alpha$ luminosities of the bright knots are on the order of $10^{38}$ erg~s$^{-1}$. We did not 
apply flux correction of possible contamination of [NII]$\lambda\lambda$6548/6583 emission for 
the H$\alpha$ photometry. Thus, the H$\alpha$ fluxes and luminosities in Table 2 are upper limits, 
possibly decreased by a few tens \%\ due to [NII] contamination. This uncertainty, however, does 
not affect the following discussion. Assuming that all the H$\alpha$ emission comes from star-
forming regions, and using the $L_{\rm H\alpha}$ to star formation rate (SFR) conversion 
relationship by \citet{Kennicutt1998}, we found that the corresponding SFRs of the knots are on 
the order of $10^{-3}$ M$_\odot$~yr$^{-1}$ (Table 2). The internal reddening correction was not 
applied in this calculation.

\begin{figure}
\epsscale{0.5}
\plotone{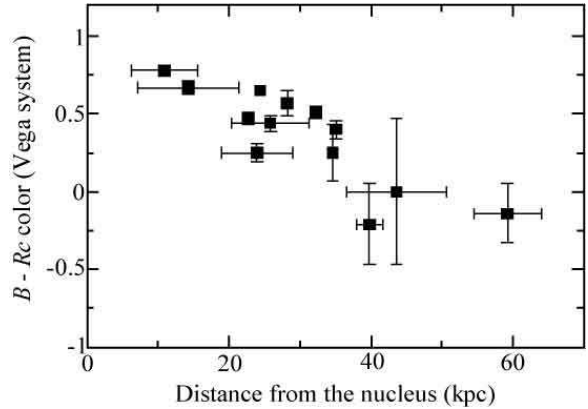}
\caption{
\label{fig10} $B - R_{\rm C}$ colors of the knots and filaments of the fireballs are shown with 
respect to the distance from the nucleus of RB199. The horizontal error bars indicate the length of 
the features.
}
\end{figure}

The fireballs are much bluer than any region of RB199 (Figure \ref{fig5}). This suggests that the 
stellar population or the dust contents of the fireballs is different from those of the galaxy disk. 
Figure \ref{fig10} shows the color distribution in the fireballs. The filaments clearly grow bluer 
with distance from the nucleus of RB199. This means either that the stellar population becomes 
younger farther out from the galaxy, or that the filaments become dustier at shorter distances.

\subsubsection{Ultraviolet data}

\begin{figure}
\epsscale{.50}
\plotone{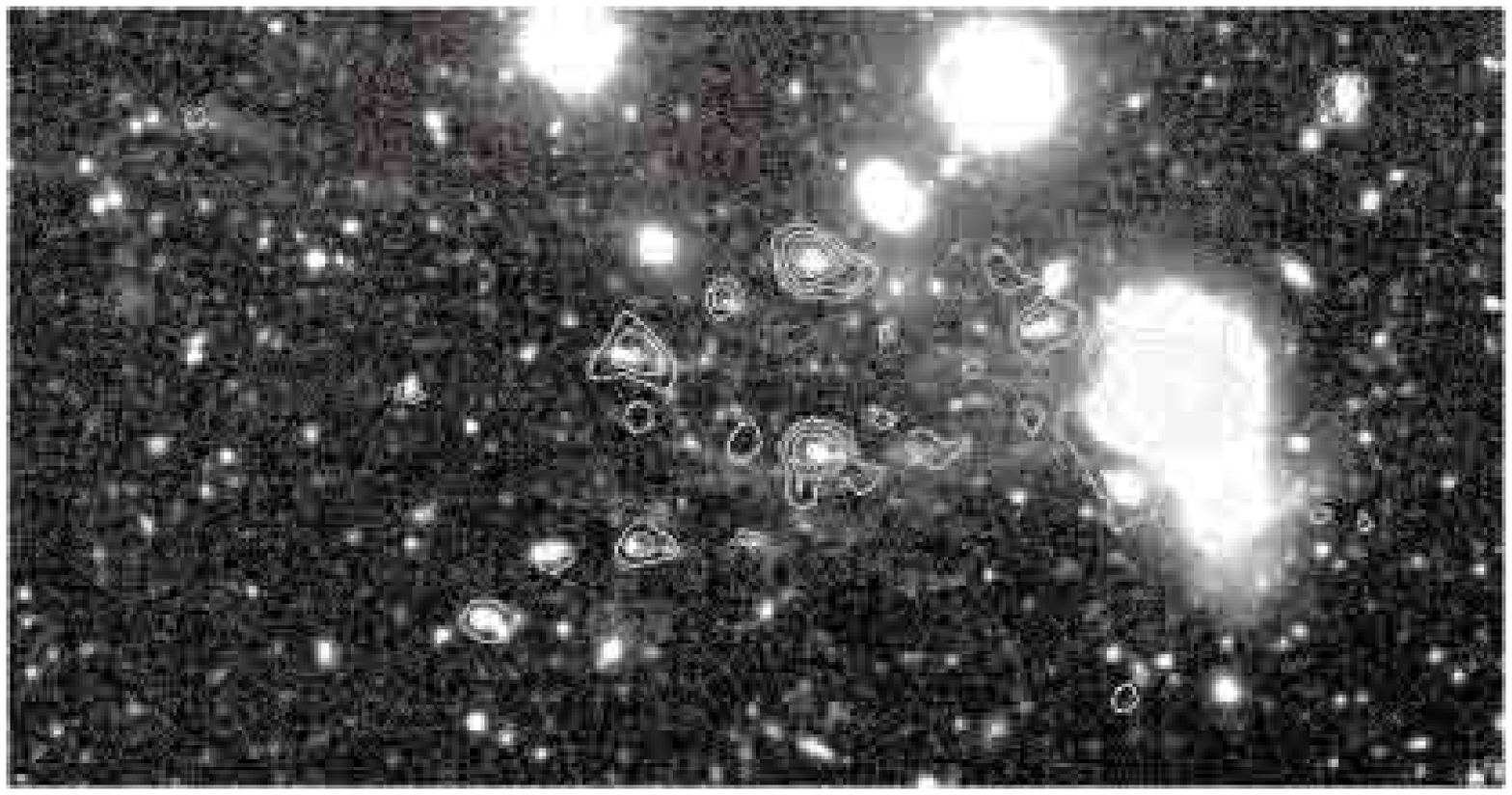}
\caption{
\label{fig11} GALEX FUV contour map overlaid on the $B$ band image of RB199.
}
\end{figure}

\begin{figure}
\epsscale{.50}
\plotone{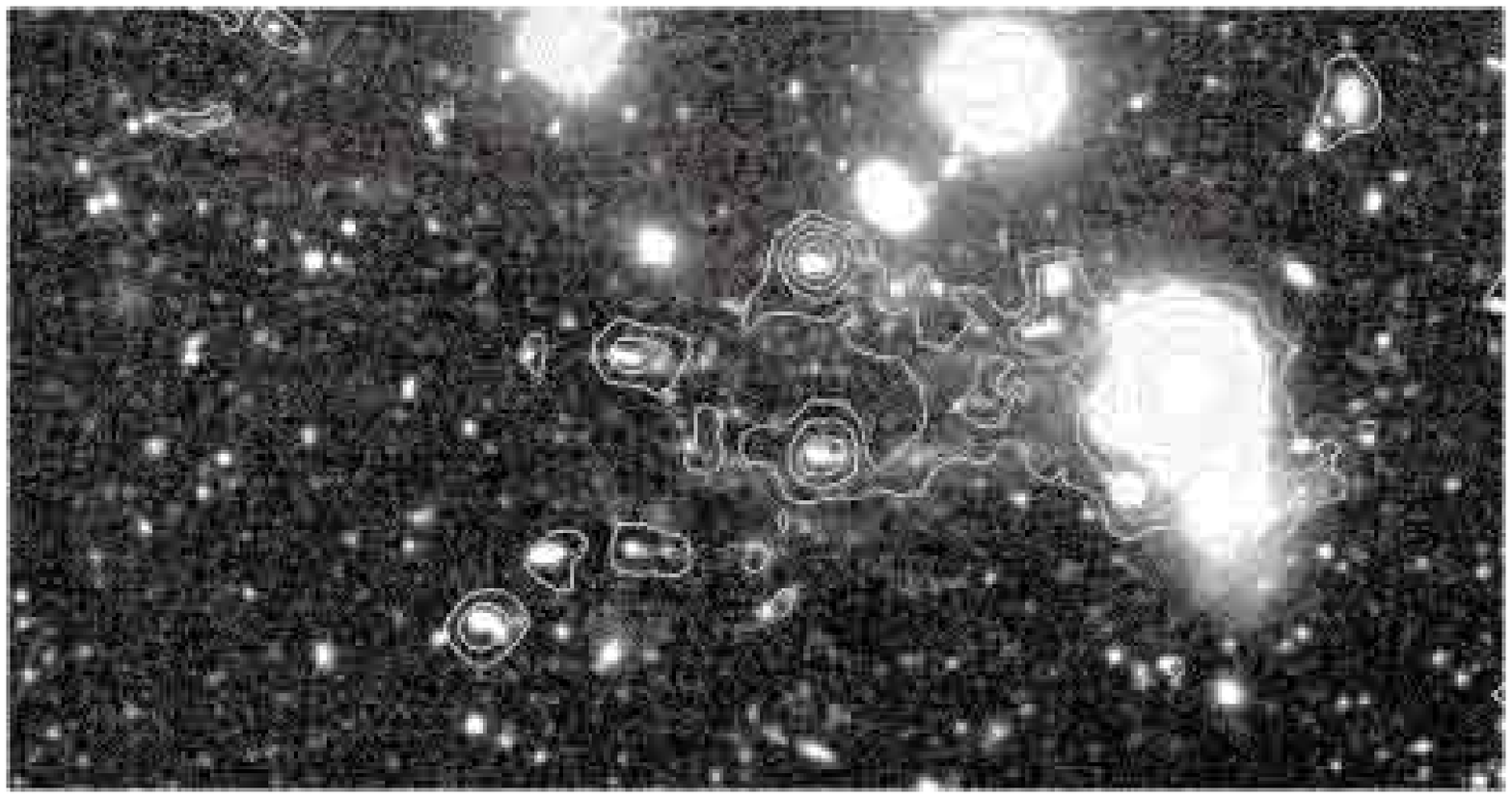}
\caption{
\label{fig12} GALEX NUV contours map overlaid on the $B$ band image of RB199.
}
\end{figure}

To investigate star formation activity in the fireballs, we downloaded UV images around RB199 
from the Web site of the second data release of Galaxy Evolution Explorer (GALEX) Nearby 
Galaxy Survey\footnote{http://galex.stsci.edu/GR2/} \citep{Martin2005}. The FUV 
($\lambda\lambda 1344 - 1786$\AA) and NUV ($\lambda\lambda 1771 - 2831$\AA) contour maps 
are superposed on the grayscale $B$ band image in Figures \ref{fig11} and \ref{fig12}.

Note that most of the bright knots in the fireballs are bright in both the FUV and NUV bands. Knots 
1--5 are clearly visible both in the FUV and NUV images.  In contrast, knot 6 can be seen only in 
the FUV image. Filaments 1 and 3 are also visible in the NUV image. In addition, very faint 
features are marginally detected in both FUV and NUV images at the position of filament 5 
(Figures \ref{fig11} and \ref{fig12}).

The UV magnitudes listed on the GALEX Web site and the ${\rm FUV} - B$ colors in the AB 
system for the bright knots are summarized in Table 3. All the bright knots are detected in the FUV 
band. Although knot 5 is visible in the NUV image (Figure \ref{fig12}), the NUV magnitude is not 
listed on the Web site. We thus measured the NUV magnitude of knot 5 directly from the NUV 
image, calibrating using the magnitude of knot 1. We also obtained the NUV magnitudes of 
filaments 1 and 3 in the NUV image (Table 3).
 
Knot 1 and 2 have FUV magnitudes of $\approx 21.5$ and the ${\rm FUV} - {\rm NUV}$ colors 
of these knots are almost flat. Knot $3 + 4$ is about one magnitude fainter than knots 1 and 2 in the 
NUV, and slightly redder. Knot 5 and 6 are almost two magnitudes fainter than knots 1 and 2 in the 
FUV band. Using the conversion formula from FUV and NUV magnitudes to SFR by 
\citet{Iglesias2006}, we derived the SFRs of knots 1--3 as $\sim 2-5 \times 10^{-
3}$ M$_\odot$~yr$^{-1}$ from the FUV magnitudes and $\sim 6-10 \times 10^{-
3}$ M$_\odot$~yr$^{-1}$ from the NUV magnitudes. These values are somewhat larger than, but 
still consistent with, the SFRs derived from H$\alpha$ luminosities (see Table 2).

Note that the ${\rm FUV} - B$ colors of the bright knots are all very blue, $-0.5- -0.9$, indicating 
that active star formation occurs in the knots. These blue colors and the strong H$\alpha$ emission 
seen at the southern sides of the bright knots strongly support the argument that the knots are 
current star-forming sites. Comparison of the optical-UV colors of the fireballs with color 
evolution models of star-forming galaxies is made in the next section.

\subsubsection{Comparison of the colors with star-formation models}

To estimate the age of the filaments, we compared their colors with model calculations of color 
evolution of star-forming stellar systems using PEGASE Ver.2 \citep{Fioc1997}. We assumed that 
the initial mass function (IMF) follows the Salpeter IMF from 0.1 M$_\odot$ to 120 M$_\odot$.  
Initial metallicity is 0.2 Z$_\odot$ in this calculation. Dust extinction was not included in our 
calculations. We derived color evolutions for four cases: constant star formation (``CS'' model), 
exponentially decaying star formation with a decay timescale of 100 Myr (``E100'' model), 200 
Myr (``E200'' model), and 300 Myr (``E300'' model).
   
\begin{figure}
\epsscale{.50}
\plotone{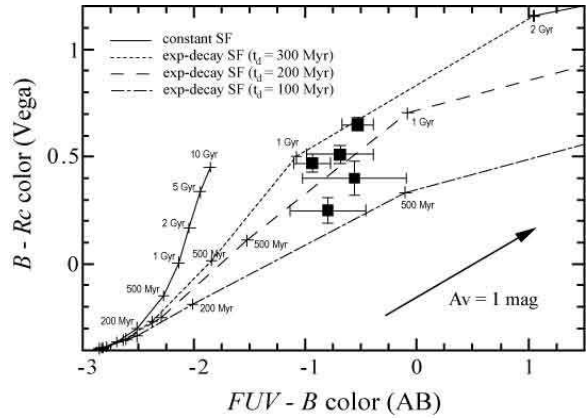}
\caption{
\label{fig13} Color--color diagram of $B - R_{\rm C}$ versus ${\rm FUV} - B$ for the bright 
knots and the blue filaments of the fireballs. The black squares represent the knot colors. The solid 
lines are the model prediction for the constant star-formation case. The dot-dashed lines, the dashed 
lines, and the dotted lines are the predictions of exponential decaying star formation models with 
decay timescales of 100 Myr (E100 model), 200 Myr (E200 model), and 300 Myr (E300 model), 
respectively. The arrows show the effect of internal dust extinction on the observed colors.
}
\end{figure}

\begin{figure}
\epsscale{.50}
\plotone{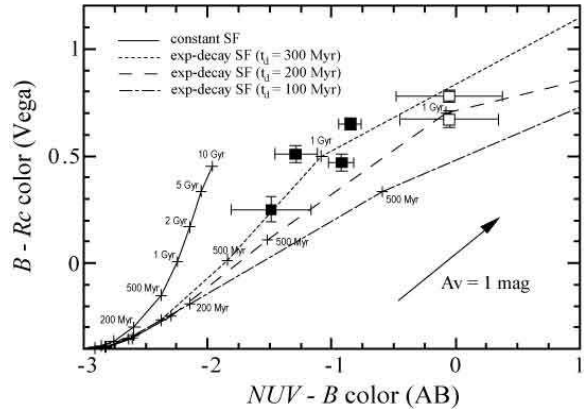}
\caption{
\label{fig14} Color--color diagram of $B - R_{\rm C}$ versus ${\rm NUV} - B$ for the fireballs. 
The white squares represent the colors of filaments 1 and 3. The other symbols and lines are the 
same as in Figure \ref{fig13}.
}
\end{figure}

Figures \ref{fig13} and \ref{fig14} show a color--color diagram of the $B - R_{\rm C}$ to ${\rm 
FUV} - B$, $B - R_{\rm C}$ to ${\rm FUV} - B$, respectively. All the colors of the bright knots 
are consistent with the E200 model or the E300 model with an age of several times $10^8$ yr to 1 
Gyr (Figures \ref{fig13} and \ref{fig14}). Figure \ref{fig14} also shows that the colors of the blue 
filaments close to the galaxy (filaments 1 and 3) are somewhat older than the bright knots, and are 
consistent with the E300 model with an age older than 1 Gyr, or the E200 model with an age of 1 
Gyr. Dust extinction trends are shown as arrows at the right-bottom corners of Figures \ref{fig13} 
and Figure \ref{fig14}. 
We derived the extinction trends using the reddening curve of star-forming galaxies proposed
by \citet{Calzetti2000}.
In both figures, the extinction trends almost follow the model loci along the 
age sequence. If dust extinction is present in the knots, their predicted ages are much younger; in 
other words, the ages estimated with Figures \ref{fig13} and \ref{fig14} are the maximum values 
for the knots. We thus conclude that the average maximum age of the bright knots is several times 
$10^8$ yr, and the maximum ages of the blue filaments that are located closer than the knots to the 
galaxy are around 1 Gyr. 

Although we have no UV data for the other filaments of the fireballs, we roughly estimate their 
ages using optical $B - R_{\rm C}$ color only. Using the E300 model, the ages of $\sim 700$ Myr, 
$\sim 1.2$ Gyr, and $\sim 500$ Myr are estimated for filament 2, filament 4, and 
H$\alpha$ filament 1, respectively, in the case of no dust extinction. The most distant blue filament, 
filament 5, is very blue, and its age is estimated to be much younger than 500 Myr. In conclusion, 
the maximum age of the filaments within 20 kpc from the galaxy is $\sim 1$ Gyr, that of the knots 
at 20--40 kpc from the galaxy is $\sim 500-1000$ Myr, and that of the farthest filaments is $\leq 
500$ Myr.

This age gradient indicates that the stellar components of the fireballs would be formed after the 
stripping event. In other words, the gas stripped by some mechanism from the disk of RB199 
travels in the intergalactic space, forming stars and leaving the formed stars along its trajectory. 
This hypothesis is supported by the spatial distribution of the H$\alpha$ emission and the bright 
knots. The bright H$\alpha$ clouds are located at the southern edges of the bright knots and
almost no H$\alpha$ emission is observed in the blue filaments. This 
indicates that active star formation is now ongoing at the far end of the blue filaments. The 
hypothesis may also be supported by the fact that the ages of the filaments and knots are roughly 
consistent with the travel time from the galaxy to the edge of the fireballs ($\sim 60-80$ kpc), 
assuming that the outflow velocity of the stripped material is on the order of $10^2$ km~s$^{-1}$. 
Although the velocities of the filaments and the knots are not known, outflow velocities of an order 
of $10^2$ km~s$^{-1}$ were observed in similar stripping phenomena 
(e.g. Yoshida et al. 2004; Cortese et al. 2006; Yagi et al. 2007).  

The color gradient in the blue filaments could also be interpreted as due to dust absorption. If this is 
the case, the above discussion of age gradient loses its meaning. However, the stellar population of 
the fireballs is certainly very young because dust absorption correction makes the true colors of the 
fireballs bluer than the observed ones (see Figures \ref{fig13} and \ref{fig14}), which strengthens 
the premise that the stars were formed very recently.

\section{Formation Mechanism of the Fireballs}

The newly found strange feature, the ``fireballs,'' around RB199 is extended in one direction from 
the galaxy as a group of narrow blue filaments and bright knots with which H$\alpha$ and UV 
emission are associated. Here we discuss possible fireball formation mechanisms.

C07 found many small blobs extending to one side of galaxies near the cores of two $z \approx 
0.2$ rich clusters in the course of a cluster survey with $Hubble$ $Space$ $Telescope$ Advanced 
Camera for Survey. \citet{Sun2007} also reported many star-forming blobs distributed around a 
very long X-ray/H$\alpha$ tail of a galaxy in the nearby rich cluster Abell~3627. The blobs 
found by C07 are distributed in faint optical filaments (C07). They detected [OII] emission around 
some of the blobs, which indicates that star formation is under way in and around the blobs. The 
morphology and colors of the blobs and the filaments found by C07 are very similar to the fireballs 
of RB199. They argued that tidal stripping by interaction between deep cluster gravitational 
potential and/or ram pressure stripping by high-speed collision between the galaxies and the 
intra-cluster medium (ICM) are responsible for forming the strange blobs they found. Following C07, we 
examine the above two processes for validity as the formation mechanism of the fireballs.

\subsection{Tidal stripping}

Tidal forces induced by galaxy--galaxy interaction disrupt the stellar disks and often throw stars far 
from the galaxies. Since RB199 appears to be a merger, it is natural to interpret the fireballs as tidal 
tails formed by tidal force induced by the merging process. The morphology of the fireballs is, 
however, very strange compared to other merger tidal tails such as NGC~4038/39 
\citep{Schweizer1978}, NGC~7252 \citep{Hibbard1996}, Arp299 \citep{Hibbard1999}, and 
IRAS~19254--7245 \citep{Mirabel1991}. In many cases, tidal tails of merging galaxies have long 
smooth morphologies extending in one or two directions from the merger main body 
\citep{Hibbard1996}. The disrupted, multiple filamentary, ``splash''-like morphology of the 
fireballs of RB199 is very rare and peculiar. Thus, from a morphological point of view, it is 
unlikely that a past merging event is responsible for forming the fireballs, although we can not rule 
out some special situation or conditions during the merger that produced such peculiar tidal features.

Tidal forces induced by interactions between RB199 and its nearby galaxies may strip the disk gas 
and stars of RB199 and may form the fireballs. In fact, several galaxies appear in the vicinity of 
RB199 (G1--G4 in Figure \ref{fig2}). Among these galaxies, G1 has a radial velocity of 6665 
km~s$^{-1}$  \citep{Colless1996}, indicating that this galaxy is a Coma member. However, no 
tidal features---bridges, tails, ripples, and so on---appear around G1. Those features are also not 
found around G2--G4. Another candidate galaxy that can strip the disk gas of RB199 is the galaxy 
about 80 kpc south of RB199 (G5 in Figure \ref{fig2}). This galaxy is located in the extension line 
of the fireballs, but the morphology of this galaxy exhibits no sign of strong tidal disturbances. To 
strip the outer disk of a galaxy through tidal forces in close encounters with other galaxies, without 
disturbing the perturbing galaxies, the perturbers must be more massive than the mass of the 
perturbed galaxy. All of the above mentioned galaxies are, however, much fainter than RB199, 
meaning that these galaxies are less massive than RB199. Furthermore, the colors of G2, G3, and 
G4 are significantly redder than the average color of Coma ellipticals, indicating that these are 
background galaxies. In fact, the photometric redshifts of these galaxies determined by the SDSS 
project are $\sim 0.3-0.4$. Therefore, it is not plausible that these galaxies significantly perturbed 
RB199 in the past.

The deep gravitational potential of a rich cluster of galaxies exerts strong tidal forces on galaxies 
crossing its core region. When a galaxy crosses a cluster core region, if the radial acceleration by 
the cluster gravity exceeds the centrifugal acceleration of the crossing galaxy, the outer part of the 
galaxy is stripped by the tidal force of the cluster (C07). The radial acceleration $a_{\rm c}$ by a 
cluster core whose mass within $r$ is $M_{\rm c}$ is written as
\begin{equation}
a_{\rm c} = 
{\rm G} M_{\rm c} \; \left[ \frac{1}{r^2} - \frac{1}{(r + R)^2} \right],
\end{equation} 
where $R$ is the radius of the crossing galaxy and $r$ is the distance between the galaxy and the 
cluster center.
The centrifugal acceleration of the galaxy is written as
\begin{equation}
a_{\rm gal} = {\rm G} \frac{M_{\rm gal}}{R^2},
\end{equation}
where $M_{\rm gal}$ is the mass of the galaxy. The projected distance from the center of the 
Coma cluster and RB199 is $\approx 0.5$ Mpc. The total (stars + gas + dark matter) mass of the 
Coma cluster within the radius of 0.5 Mpc was estimated as $\approx 3 \times 
10^{14}$ M$_\odot$  \citep{Lokas2003}. The $R$ band luminosity of RB199 is $1.4 \times 
10^9$ L$_\odot$ \citep{Mob2001}. Thus, assuming $M$/$L_R \sim 2 - 3$, which is a typical 
value of disk $M/L$ ratios for nearby spiral galaxies 
\citep{Forbes1992,Palunas2000,Yoshino2008}, the mass of the disk of RB199 is calculated as $3-
4 \times 10^9$ M$_\odot$. The roots of the blue filaments are all connected to the central ellipse of 
RB199; thus, the effective radius of the disturbed region of the galaxy is that of the ellipse, i.e., $R 
\sim 5$ kpc. With these values, one can surmise that $a_{\rm gal}$ is one order of magnitude 
larger than $a_{\rm c}$. This means that the cluster potential can hardly strip the outer disk 
component of RB199. Even if the $M$/$L_R$ of the galaxy is almost unity, $a_{\rm gal}$ is still 
larger than $a_{\rm c}$. Therefore, tidal stripping by cluster--galaxy interaction is not a possible 
candidate for a formation mechanism of the fireballs around RB199.

\subsection{Ram pressure stripping}

When a gas-rich disk galaxy collides with the hot ICM at high speed ($v\sim 10^3$ km s$^{-1}$), 
ram pressure from the ICM strips massive amounts of gas from the galaxy. This process, ram 
pressure stripping, has been discussed in many studies as an efficient gas removing mechanism for 
galaxies in clusters 
\citep{Gunn1972,Abadi1999,Fujita1999,Vollmer2001,Schulz2001,Bekki2003,Roediger2007}. 
From an observational point of view, \citet{Cayatte1990,Cayatte1994} found that HI gas is 
significantly deficient for spiral galaxies in the core region of the Virgo cluster. 
\citet{Bravo2000,Bravo2001} found strong HI deficiency for bright spiral galaxies inside a radius 
of 0.6 Mpc from the center of the Coma cluster. In addition, detailed case studies have been carried 
out for many objects through optical and radio observations (NGC~4388: Yoshida et al. 2002, 
2004; Vollmer \& Huchtmeier 2003; Oosterloo \& van Gorkom 2005; NGC~4402: Crowl et al. 
2005; NGC~4438: Chemin et al. 2005; NGC~4522: Kenny \& Koopmann 1999; Vollmer et al. 
2004).

The position and the velocity of RB199 suggest that the galaxy suffers strong ram pressure from the 
ICM through high-speed collision with it. RB199 is located at the edge of the dense ICM associated 
with the main body of the Coma cluster \citep{Poggianti2004}. In addition, the radial velocity of 
RB199 ($v \approx 8700$ km~s$^{-1}$; Mobasher et al. 2001) is quite large with respect to the 
mean radial velocity of the Coma cluster ($v \approx 6500$ km~s$^{-1}$). To determine whether 
the disk gas of RB199 can be stripped, we try to estimate the ram pressure stripping radius $R_{\rm 
strip}$ for the RB199 case. When the distribution of stars and gas of the galaxy disk is a double 
exponential, $R_{\rm strip}$ is estimated as
\begin{equation}
R_{\rm strip} = 0.5  R_{\rm 0} \times ln \left[ \frac{{\rm G} M_{\rm star} M_{\rm gas}}{v^2\; 
\rho_{\rm ICM}\; 2 \pi R_{\rm 0}^4} \right],
\end{equation}
where $M_{\rm star}$ and $M_{\rm gas}$ are the mass of the stars and the gas in the galaxy disk, 
respectively, $v$ is the galaxy velocity relative to the ICM, $\rho_{\rm ICM}$ is the density of the 
ICM, and $R_{\rm 0}$ is the radial scale length of the disk \citep{Domainko2006}. Substituting 
$4.2 \times 10^9$ M$_\odot$, $\sim 10^8$ M$_\odot$, $\sim 10^{-4}$ cm$^{-3}$ and $\sim 
2000$ km~s$^{-1}$ for $M_{\rm star}$, $M_{\rm gas}$, $\rho_{\rm ICM}$ and $v$, 
respectively, we obtain
\begin{equation}
R_{\rm strip} = 0.5  \left( \frac{R_{\rm 0}}{{\rm kpc}} \right) \times ln \left[ 25.4 \times 
\left( \frac{R_{\rm 0}}{{\rm kpc}} \right)^{-4} \right].
\end{equation}
We fitted an exponential disk function to the light profile of the central ellipse of RB199 along 
PA$ = $70$^{\circ}$ and found that $R_{\rm 0} \approx 1.5$ kpc. Note that RB199 has an 
irregular shape and does not have a clear exponential disk. Hence the $R_{\rm 0}$ we derived 
above is not a correct ``effective radius.'' This $R_{\rm 0}$ is, however, an indicator of the surface 
density distribution of the galaxy, and thus is useful for determining whether the disk gas of RB199 
can be stripped by the ram pressure of the ICM. With this value, we derived $R_{\rm strip} \approx 
1.2$ kpc. This means that almost all the gas in the disk of RB199 would be stripped by the
ram pressure of the ICM.

The morphology of the fireballs also supports the ram pressure stripping hypothesis. The fireballs 
extend from one side of RB199, and ram pressure stripping forms a filamentary structure extending 
on one side from the colliding galaxy 
(Abadi et al. 1999; Quilis, Moor \& Bower 2000; Roediger \& Hensler 2005;
Kronberger et al. 2008). In 
addition, the configuration in which the ionized gas is far more extended than the stars (the blue 
filaments and the bright knots) is naturally explained by ram pressure. The gas clouds undergo the 
ram pressure of the ICM and are accelerated by it continuously after the stripping, whereas the stars 
are not affected by the ram pressure and suffer the gravitational force of the galaxy. Hence the gas 
clouds stretch far away from the galaxy while the stars remain relatively nearby. Therefore, we 
propose that ram pressure stripping is the primary formation mechanism of the fireballs.

A major argument against the ram pressure stripping hypothesis is the fact that the blue filaments 
and the bright knots of the fireballs consist mainly of stars. Ram pressure stripping can remove only 
gas from a galaxy and cannot greatly affect the stellar system of the galaxy. Stars are too massive, 
and their collisional cross sections are too small, to respond to drag force from the ram pressure of 
the ICM flow. However, note that the stellar population of the fireballs is very young and must 
largely have been formed after stripping from the host galaxy. In other words, only the gas in the 
disk was stripped in the initial phase, and the stars were then formed in the stripped gas.

Sun et al. (2007) also argued that the blue blobs they found may have been formed in the stripped 
gas by the ram pressure. Recently, \citet{Kron2008} performed numerical simulations focusing on 
star formation due to ram pressure stripping and found that strong star formation is triggered in the 
central region of their model galaxy as well as in the stripped gas filaments behind the galaxy. The 
star formation blobs, whose sizes and masses are of an order of 1 kpc and 10$^7$ M$_\odot$, 
respectively, extended up to 100 kpc from the galaxy \citep{Kron2008}. In their simulations, the 
stripped gas filaments have narrow and slightly wiggled straight shapes, whereas the newly formed 
stars form somewhat diffuse wide filaments. Note that the mass and size of the blobs and the 
morphology of the complex of the blobs and the filaments derived in the simulations of 
\citet{Kron2008} strongly resemble those of the fireballs we found around RB199. Accordingly, 
Kronberger et al.'s results strongly support the hypothesis that the fireballs are formed by ram 
pressure stripping.

In the case that ram pressure stripping is the primary fireball formation driver, the stripped neutral 
hydrogen (HI) gas would be observed around RB199. Bravo-Alfaro et al. (2000, 2001) made deep 
HI observations of the Coma cluster using the VLA. RB199 and its surrounding area were within 
the observation areas of fields 7 and 10 in their study \citep{Bravo2000}. They observed no sign of 
HI gas around RB199. The primary cause for this is that the recession velocity of RB199 ($\approx 
8700$ km s$^{-1}$) is far outside the frequency bands (6.25 MHz) of their observations, which 
were centered at radial velocities of  5500 km s$^{-1}$ (``field 7'') and 7300 km s$^{-1}$ (``field 
10'') \citep{Bravo2000}.

\subsection{Ram pressure stripping with galaxy--galaxy merger}

The morphology of RB199 indicates that this galaxy is a merger remnant. We are thus led to the 
idea that a galaxy--galaxy merger in addition to ram pressure of the ICM may have played an 
important role in efficiently stripping the gas from RB199. A galaxy--galaxy merger perturbs the 
interstellar gas in the galaxy disk significantly, and some portion of the gas would flare up above 
the disk. A combination of this effect and the ram pressure may make gas stripping from the galaxy 
very efficient. \citet{Struck2004} performed numerical simulations of ram pressure stripping for 
merging galaxies infalling to a cluster. Their results support that the above simple picture is correct 
in principle. They found that strongly interacting galaxy pairs tend to be stripped of much more gas 
than isolated galaxies by the ram pressure of the ICM \citep{Struck2004}. This process may be 
very efficient in stripping the interstellar medium from a galaxy disk, and may strip heavy elements 
other than HI gas, such as metals, molecules, or dust that make star formation in the stripped gas 
possible.

One of the most striking features formed by a combination of galaxy--galaxy interaction and ram 
pressure stripping ever found is the Blue Infalling Group (BIG) in Abell 1367 
\citep{Sakai2002,Gavazze2003,Cortese2006}. Very extended ionized gas filaments and blue tails 
are observed in BIG \citep{Cortese2006}. In addition, many star-forming dwarf galaxies are 
distributed in BIG \citep{Sakai2002}. \citet{Cortese2006} concluded that the combined action of 
tidal forces among the galaxies and ram pressure by the ICM produced that peculiar structure. 
Although BIG is much larger and more luminous than the fireballs around RB199, the phenomena 
appear similar. In addition, the host galaxies of the blobs of C07 and Sun et al. (2007) have 
disturbed morphologies, suggesting a past merger. These observational results support the idea that 
a galaxy--galaxy interaction/merger in addition to ram pressure stripping may make a significant 
contribution to the formation of fireballs and similar objects.

In conclusion, ram pressure stripping is the key driver in fireball formation. A galaxy--galaxy 
merger in addition to the ram pressure may make a considerable contribution to their formation. 
Tidal interactions between RB199 and the gravitational potential of the Coma cluster would not be 
strong enough to play a major role in formation of the fireballs.

\section{Summary}

We found a strange complex (the ``fireballs'') of blue filaments and H$\alpha$ clouds that extends 
up to 80 kpc toward the south from the merging galaxy RB199 in the Coma cluster. The narrow 
blue filaments are extended in straight shapes and several bright blue knots are located at the south 
end of the blue filaments. Strong H$\alpha$ emission is associated with the knots. In addition, faint, 
narrow H$\alpha$-emitting filaments extend farther from the southern edge of the knots. 

The total $B$ and $R_{\rm C}$ magnitudes of the fireballs are 19.87 and 19.33, respectively. The 
luminosities of the H$\alpha$ emission associated with the bright knots are $ 1-6 \times 
10^{38}$ erg s$^{-1}$, which correspond to a star-formation rate of $\sim 10^{-
3}$ M$_\odot$ yr$^{-1}$, assuming all the H$\alpha$ comes from star-forming activity. The 
average color of the fireballs is $B - R_{\rm C}\approx 0.5$, which is bluer than RB199 ($B - 
R_{\rm C} = 0.99$), and the blue filaments and the bright knots grow bluer with distance from the 
nucleus of the galaxy. These observational facts suggest that the stars of which the fireballs are 
composed were formed within several times $10^8$ yr and that the stellar population grows 
younger farther from the galaxy.

Almost no H$\alpha$ emission is observed in the blue filaments but strong H$\alpha$ and UV emissions 
are associated with the bright knots. These characteristics indicate that star formation has already 
ceased in the blue filaments and is now in progress in the bright knots. The gas stripped by some 
mechanism from the disk of RB199 may be traveling in the intergalactic space, forming stars and 
leaving the formed stars along its trajectories. The $R_{\rm C}$-band absolute magnitudes, the half 
light radii, and the estimated masses of bright knots in the fireballs are $\sim -12 - -13$ mag, $\sim 
200-300$ pc and  $\sim 10^{6-7}$ M$_{\odot}$, respectively. These values are similar to those of 
faint dwarf spheroidals in the Virgo cluster or the faint end of local dwarf galaxies. This suggests 
that the bright knots are gravitationally self-bound systems.

The most plausible formation mechanism of the fireballs is ram pressure stripping by the hot ICM. 
Tidal stripping by galaxy--galaxy interaction or galaxy--cluster interaction would not be a primary 
mechanism for the formation of the fireballs. A galaxy--galaxy merger, however, is probably 
important to make subsequent ram pressure stripping of the ICM more effective in forming this 
strange structure.

Although whether the stars in the fireballs are all gravitationally bound to RB199 is not known, 
some structures close to the galaxy may fall back to the galaxy in the future, oscillating between the 
two sides of the galaxy disk and eventually forming the halo star population of RB199.
Some part of the fireballs may be stripped from RB199 in the future by tidal interaction with 
nearby galaxies or the cluster gravitational potential, even if they are currently bound to the galaxy. 
If this is the case, the stripped stars may form a diffuse intra-cluster population 
\citep{Gregg1998,Okamura2002,Murante2004,Krick2006,Gerhard2007}.

The fireballs we found in the Coma cluster resemble the compact young blobs found by C07 and 
Sun et al. (2007) around some cluster galaxies. C07 argued that formation of these compact blobs is 
rather rare in clusters at $z \approx 0.2$. They found only two examples out of 130 cluster spiral 
galaxies at $z \approx 0.2$. They also predicted that these features would be more remarkable at $z 
\geq 0.2$ because the galaxy infalling rate and the gas content of each galaxy are higher in high-
$z$ clusters than in local clusters. However, our findings show that formation of compact stellar 
systems like C07's blobs around cluster galaxies is still ongoing in the local universe. The Coma 
cluster has a large substructure of the ICM and a group of galaxies associated with this substructure, 
indicating that this cluster itself is a merging system \citep{Colless1996}. RB199 is located near 
the border between the main body and the substructure of the ICM \citep{Poggianti2004}. The 
special conditions of the Coma cluster and RB199 may produce a special environment around 
RB199,namely, high-ensity ambient gas, high-peed infall, and/or frequent galaxy--galaxy 
interactions

\acknowledgments

We thank the staff of the Subaru telescope for their kind help with the observations.
This work was carried out in part using data obtained by a collaborative study on the Coma cluster. 
We thank our collaborators in this study, David Carter, Alister Graham, Shardha Jogee, Neal Miller, 
and Bahram Mobasher, for their encouragement and helpful comments. We also thank Ikuru Iwata 
for his help in calculating the color evolution model of a star-forming galaxy.
We are grateful to the anonymous referee for constructive suggestions that helped to greatly 
improve the manuscript. This study was in part carried out using the facilities at the Astronomy 
Data Center (ADC), National Astronomical Observatory of Japan. This research made 
use of NASA's Astrophysics Data System Abstract Service, NASA/IPAC Extragalactic Database, 
and GALEX Release 2 database. We thank the National Institute of Information and 
Communications Technology for their support on the high-speed network connection for data 
transfer and analysis. This work was financially supported in part by a Grant-in-Aid for Scientific 
Research (B) No.18340055 from the Japan Society for the Promotion of Science
(JSPS).



\clearpage

\begin{deluxetable}{lcccc}
  \tabletypesize{\footnotesize}
  \tablewidth{0pt}
  \footnotesize
  \tablecaption{Observation Log}
  \tablehead{
    \colhead{filter} & 
    \colhead{PSF size} &
    \colhead{SB$_{\rm lim}$(ABmag arcsec$^{-2}$)\tablenotemark{a}} &
    \colhead{date (UT)}&
    \colhead{exposure}
  }
  \startdata
$B$ & 1\arcsec.06 & 28.8 & 2006-04-28 &  2 $\times$ 450sec  \\
    &        &      & 2006-05-03 &  3 $\times$ 450sec  \\
    &        &      & 2007-05-13 &  5 $\times$ 600sec  \\
$R_{\rm C}$ & 0\arcsec.81 & 28.0 & 2006-04-28 & 11 $\times$ 300 sec  \\
    &        &      & 2007-05-14 &  5 $\times$ 360 sec  \\
    &        &      & 2007-05-15 &  1 $\times$ 300 sec  \\
NB  & 0\arcsec.82 & 27.8 & 2006-04-28 &  5 $\times$ 1800 sec \\
    &        &      & 2007-05-15 &  8 $\times$ 1800 sec \\
\enddata

\tablenotetext{a}{
SB$_{\rm lim}$ represent 1-$\sigma$ fluctuation of 
a circular aperture of 2 arcsec diameter. The 
$R_{\rm C}$ and NB are calibrated by short exposures taken on 
2006-05-03, but the mosaic 
image does not include them.}
\end{deluxetable}


\begin{deluxetable}{lcccccccc}
  \tabletypesize{\footnotesize}
  \tablewidth{0pt}
  \footnotesize
  \tablecaption{Photometry of the Fireballs around RB199}
  \tablehead{
    \colhead{ID} & 
    \colhead{distance\tablenotemark{a}} &
    \colhead{size\tablenotemark{b}} &
    \colhead{$m_{B}$} &
    \colhead{$m_{R_{\rm C}}$} &
    \colhead{$B - R_{\rm C}$} & 
    \colhead{$M_{\rm L}$\tablenotemark{c}} &
    \colhead{$L_{\rm H\alpha}$ \tablenotemark{d}} &
    \colhead{SFR \tablenotemark{e}}
  }
  
  \startdata 
    knot 1            & 23 & 2.6$\times$0.9 & 22.52$\pm 0.03$ & 22.04$\pm 0.03$ & 0.47$\pm 
0.04$  & 8.8  & 34   & 27  \\
    knot 2            & 24 & 1.8$\times$1.0 & 22.18$\pm 0.02$ & 21.53$\pm 0.02$ & 0.65$\pm 
0.03$  & 14   & 59   & 47  \\
    knot 3            & 32 & 1.9$\times$0.6 & 23.19$\pm 0.03$ & 22.68$\pm 0.03$ & 0.51$\pm 
0.04$  & 4.8  & 9.9  & 7.8 \\
    knot 4            & 34 & 0.7$\times$0.7 & 25.13$\pm 0.12$ & 24.88$\pm 0.13$ & 0.25$\pm 
0.18$  & 0.76 & 2.1  & 1.7 \\
    knot 5            & 35 & 0.7$\times$0.7 & 24.06$\pm 0.04$ & 23.66$\pm 0.04$ & 0.40$\pm 
0.06$  & 2.3  & 30   & 23  \\
    knot 6            & 28 & 0.6$\times$1.6 & 24.23$\pm 0.06$ & 23.66$\pm 0.05$ & 0.57$\pm 
0.08$  & 2.0  & 3.2  & 2.5 \\
    filament 1        & 14 & 14 & 22.03$\pm 0.03$ & 21.36$\pm 0.03$ & 0.67$\pm 0.04$  & 16   &  
-   & -   \\
    filament 2        & 24 &  9 & 22.53$\pm 0.04$ & 22.28$\pm 0.03$ & 0.25$\pm 0.04$  & 6.7  & 
0.9  & 0.7 \\
    filament 3        & 11 & 10 & 22.10$\pm 0.02$ & 21.32$\pm 0.05$ & 0.78$\pm 0.06$  & 16   &  
-   &  -  \\
    filament 4        & 26 &  5 & 22.83$\pm 0.04$ & 22.39$\pm 0.03$ & 0.44$\pm 0.05$  & 6.4  & 
5.9  & 4.7 \\
    filament 5        & 59 &  8 & 23.22$\pm 0.15$ & 23.36$\pm 0.11$ & -0.14$\pm 0.16$ & 2.5  & 
3.8  & 3.0 \\
    H$\alpha$ filament 1 & 44 & 14 & 25.39$\pm 0.48$ & 25.39$\pm 0.32$ & 0.00$\pm 0.48$  & 
0.68  & 28   & 23  \\
    H$\alpha$ filament 2 & 40 &  4 & 24.86$\pm 0.12$ & 25.07$\pm 0.23$ & -0.21$\pm 0.26$  & 
0.52 & 12   & 9.2 \\
    H$\alpha$ cloud 1 & 80 & - & 26.04$\pm 0.25$ &   -   &   -   & 0.62\tablenotemark{f} & 9.0  & 
7.2 \\
    H$\alpha$ cloud 2 & 79 & - & 26.29$\pm 0.62$ &   -   &   -   & 0.50\tablenotemark{f} & 11   & 
8.6 \\
    \hline
    \\
    Total             & 83\tablenotemark{g} &  & 19.87 & 19.33 & 0.54  & 110  & 225  & 179 \\
  \enddata

  \tablenotetext{a}{distance from the nucleus of RB199 in units of kpc.}
  \tablenotetext{b}{sizes of the knots, filaments, and cloud in units of kpc. The sizes are measured
at the $B$ band surface brightness of 28 mag~arcsec$^{-2}$.} 
  \tablenotetext{c}{mass derived from $R_{\rm C}$ band magnitude in units of 
$10^6$ M$_\odot$ assuming that $M/L_R = 1$.}
  \tablenotetext{d}{H$\alpha$ luminosity in unit of $10^{37}$ ergs s$^{-1}$.
Contamination of [NII]$\lambda\lambda 6548/6583$ emission and internal dust extinction
is not corrected.} 
  \tablenotetext{e}{star-formation rate calculated from $L_{\rm H\alpha}$ in units of $10^{-
4}$ M$_\odot$ yr$^{-1}$ assuming that all the H$\alpha$ emission is produced by star-formation 
activity.}
  \tablenotetext{f}{derived from $B$ band magnitude assuming that $M/L_B = 1$.}
  \tablenotetext{g}{distance between the edge of the H$\alpha$ cloud 1 and the nucleus of 
RB199.}

\end{deluxetable}


\begin{deluxetable}{lccccc}
  \tabletypesize{\footnotesize}
  \tablewidth{0pt}
  \footnotesize
  \tablecaption{Ultraviolet Magnitudes and Optical-UV Colors of the Fireballs}
  \tablehead{
    \colhead{ID} & 
    \colhead{GALEX ID} &
    \colhead{GALEX FUV\tablenotemark{a}} &
    \colhead{GALEX NUV\tablenotemark{a}} &
    \colhead{${\rm FUV} - B$} &
    \colhead{${\rm NUV} - B$}\\
    \colhead{} &
    \colhead{} &
    \colhead{(AB mag)} &
    \colhead{(AB mag)} &
    \colhead{(AB mag)} &
    \colhead{(AB mag)}
  }
  \startdata
    knot 1    & J125841.9 + 274451 & 21.44$\pm 0.16$ & 21.46$\pm 0.09$ & -0.94$\pm 0.16$ & -
0.92$\pm 0.10$ \\
    knot 2    & J125843.8 + 274450 & 21.51$\pm 0.14$ & 21.19$\pm 0.08$ & -0.53$\pm 0.14$ & -
0.85$\pm 0.09$ \\
    knot 3 + 4  & J125842.9 + 274427 & 22.36$\pm 0.30$ & 21.76$\pm 0.17$ & -0.69$\pm 0.30$ & 
-1.29$\pm 0.21$ \\
    knot 5    & J125841.0 + 274429 & 23.12$\pm 0.34$ & 22.43$\pm 0.33$\tablenotemark{b} & -
0.80$\pm 0.34$ & -1.49$\pm 0.33$ \\
    knot 6    & J125843.4 + 274439 & 23.53$\pm 0.45$ &   -             & -0.56$\pm 0.46$ &   - \\
    filament 1 & -               & -               & 21.84$\pm 0.40$\tablenotemark{b} & -               & -
0.19$\pm 0.40$ \\
    filament 3 & -               & -               & 21.91$\pm 0.43$\tablenotemark{b} & -               & -
0.19$\pm 0.43$ \\
  \enddata

  \tablenotetext{a}{Taken from the Web site of GALEX data release 2 (DR2), except for the NUV 
magnitudes of knot 5, filament 1, and filament 3.}
  \tablenotetext{b}{Directly measured in the NUV image calibrated by the GALEX DR2 NUV 
magnitude of knot 1.} 

\end{deluxetable}


\begin{thebibliography}{}

\bibitem[Abadi, Moore \& Bower(1999)]{Abadi1999}
Abadi, M. G., Moore, B., \& Bower, R. G. 1999, \mnras, 308, 947

\bibitem[Adami et al.(2007)]{Adami2007}
Adami, C., et al. 2007, \aap, 472, 749

\bibitem[Bekki \& Couch(2003)]{Bekki2003}
Bekki, K., \& Couch, W. J. 2003, \apjl, 596, L13

\bibitem[Bekki, Couch \& Shioya(2002)]{Bekki2002}
Bekki, K., Couch, W. J., \& Shioya, Y. 2002, \apj, 577, 651

\bibitem[Belokurov et al.(2007)]{Belokurov2007}
Belokurov, V., et al. 2007, \apj, 654, 897

\bibitem[Boselli \& Gavazzi(2006)]{Boselli2006}
Boselli, A., \& Gavazzi, G. 2006, \pasp, 118, 517

\bibitem[Bravo-Alfaro et al.(2000)]{Bravo2000}
Bravo-Alfaro, H., Cayatte, V., van Gorkom, J. H., \& Balkowski, C. 2000, \aj, 119, 580

\bibitem[Bravo-Alfaro et al.(2001)]{Bravo2001}
Bravo-Alfaro, H., Cayatte, V., van Gorkom, J. H., \& Balkowski, C. 2001, \aap, 379, 347

\bibitem[Butcher \& Oemler(1978)]{Butcher1978}
Butcher, H., \& Oemler, A. 1978, \apj, 219, 18

\bibitem[Butcher \& Oemler(1984)]{Butcher1984}
Butcher, H., \& Oemler, A. 1984, \apj, 285, 426

\bibitem[Calzetti et al.(2000)]{Calzetti2000}
Calzetti, D. et al. 2000, \apj, 533, 682

\bibitem[Cayatte et al.(1990)]{Cayatte1990}
Cayatte, V., van Gorkom, J. H., Balkowski, C., \& Kotanyi, C. 1990, \aj, 100, 604

\bibitem[Cayatte et al.(1994)]{Cayatte1994}
Cayatte, V., Kotanyi, C., Balkowski, C., \& van Gorkom, J. H. 1994, \aj, 107, 1003

\bibitem[Chemin et al.(2005)]{Chemin2005}
Chemin, L., et al. 2005, \aap, 436, 469

\bibitem[Colless \& Dunn(1996)]{Colless1996}
Colless, M., \& Dunn, A. M. 1996, \apj, 458, 435

\bibitem[Conselice \& Gallagher(1998)]{Conselice1998}
Conselice, C. J., \& Gallagher, J. S. 1998, \mnras, 297, L34

\bibitem[Cortese et al.(2006)]{Cortese2006}
Cortese, L., et al. 2006, \aap, 453, 847

\bibitem[Cortese et al.(2007)]{Cortese2007}
Cortese, L., et al. 2007, \mnras, 376, 157 (C07)

\bibitem[Crowl et al.(2005)]{Crowl2005}
Crowl, H. H., Kenney, J. D. P., van Gorkom, J. H., \& Vollmer, B. 2005, \apj, 130, 65

\bibitem[Domainko et al.(2006)]{Domainko2006}
Domainko, W., et al. 2006, \aap, 452, 795

\bibitem[Dressler(1980)]{Dressler1980}
Dressler, A. 1980, \apj, 236, 351

\bibitem[Dressler(1994)]{Dressler1994}
Dressler, A. 1994, \apj, 430, 107

\bibitem[Dressler et al.(1997)]{Dressler1997}
Dressler, A., et al. 1997, \apj, 490, 577

\bibitem[Drinkwater et al.(2004)]{Drinkwater2004}
Drinkwater, M. J., et al. 2004, PASA, 21, 375

\bibitem[Evstigneeva et al.(2007)]{Evs2007}
Evstigneeva, E. A., Gregg, M. D., Drinkwater, M. J., \& Hilker, M. 2007, \aj, 133, 1722

\bibitem[Farouki \& Shapiro(1981)]{Farouki1981}
Farouki, R., \& Shapiro, S. L. 1981, \apj, 243, 32 

\bibitem[Fioc \& Rocca-Volmerange(1997)]{Fioc1997}
Fioc, M., \& Rocca-Volmerange, B. 1997, \aap, 326, 950

\bibitem[Forbes(1992)]{Forbes1992}
Forbes, D. A. 1992, \aaps, 92, 583

\bibitem[Fujita \& Nagashima(1999)]{Fujita1999}
Fujita, Y., \& Nagashima, M. 1999, \apj, 516, 619.

\bibitem[Gavazzi et al.(2003)]{Gavazze2003}
Gavazzi, G., et al. 2003, \apj, 597, 210

\bibitem[Gerhard et al.(2007)]{Gerhard2007}
Gerhard, O., et al. 2007, \aap, 468, 815-822

\bibitem[Goto et al.(2003)]{Goto2003}
Goto, T., et al. 2003, \mnras, 346, 601

\bibitem[Gregg \& West(1998)]{Gregg1998}
Gregg, M. D., \& West, M. J. 1998, \nat, 396, 549

\bibitem[Gunn \& Gott(1972)]{Gunn1972}
Gunn, J. E., \& Gott, J. R. 1972, \apj, 176, 1

\bibitem[Harsono \& De Propris(2007)]{Harsono2007}
Harsono, D., \& De Propris, R. 2007, \mnras, 380, 1036

\bibitem[Henriksen \& Byrd(1996)]{Henriksen1996}
Henriksen, M., \& Byrd, G. 1996, \apj, 459, 82

\bibitem[Hibbard \& van Gorkom(1996)]{Hibbard1996}
Hibbard, J. E., \& van Gorkom, J. H. 1996, \aj, 111, 655

\bibitem[Hibbard \& Yun(1999)]{Hibbard1999}
Hibbard, J. E., \& Yun, M. S. 1999, \aj, 118, 162

\bibitem[Hilker et al.(2007)]{Hilker2007}
Hilker, M., et al. 2007, \aap, 463, 119

\bibitem[Icke(1985)]{Icke1985}
Icke, V. 1985, \aap, 144, 115

\bibitem[Iglesias-Paramo et al.(2006)]{Iglesias2006}
Iglesias-Paramo, J., et al. 2006, \apjs, 164, 38

\bibitem[J$\acute{\rm a}$chym et al.(2007)]{Jac2007}
J$\acute{\rm a}$chym, P, et al. 2007, \aap, 472, 5

\bibitem[Kennicutt(1998)]{Kennicutt1998}
Kennicutt, R. C. 1998, \apj, 498, 541

\bibitem[Kenny \& Koopmann(1999)]{Kenny1999}
Kenny, J. D. P., \& Koopmann, R. A. 1999, \aj, 117, 181

\bibitem[Krick, Bernstein \& Pimbblet(2006)]{Krick2006}
Krick, J. E., Bernstein, R. A., \& Pimbblet, K. A. 2006, \aj, 131, 168

\bibitem[Kronberger et al.(2008)]{Kron2008}
Kronberger, T., Kapferer, W., Ferrari, C., Unterguggenberger, S., \& Schindler, S. 2008, \aap, 481, 
337

\bibitem[Larson, Tinsley \& Caldwell(1980)]{Larson1980}
Larson, R. B., Tinsley, B. M., \& Caldwell, C. N. 1980, \apj, 237, 692

\bibitem[Lokas \& Mamon(2003)]{Lokas2003}
Lokas, E. L., \& Mamon, G. A. 2003, \mnras, 343, 401

\bibitem[Martin et al.(2005)]{Martin2005}
Martin, D. C., et al. 2005, \apj, 619, L1

\bibitem[Milne et al.(2007)]{Milne2007}
Milne, M. L., et al. 2007, \aj, 133, 177

\bibitem[Mirabel, Lutz \& Maza(1991)]{Mirabel1991}
Mirabel, I. F., Lutz, D., \& Maza, J. 1991, \aap, 243, 367

\bibitem[Miyazaki et al.(2002)]{Miyazaki2002}
Miyazaki, S., et al. 2002, \pasj, 54, 833

\bibitem[Mobasher et al.(2001)]{Mob2001}
Mobasher, B., et al. 2001, \apjs, 137, 279

\bibitem[Moore et al.(1996)]{Moore1996}
Moore, B., Katz, N., Lake, G., Dressler, A., \& Oemler, A. 1996, \nat, 379, 613

\bibitem[Murante et al.(2004)]{Murante2004}
Murante, G., et al. 2004, \apjl, 607, L83

\bibitem[Nulsen(1982)]{Nulsen1982}
Nulsen, P. E. J. 1982, \mnras, 198, 1007

\bibitem[Okamura et al.(2002)]{Okamura2002}
Okamura, S., et al. 2002, \pasj, 54, 883

\bibitem[Oosterloo \& van Gorkom(2005)]{Oosterloo2005}
Oosterloo, T., \& van Gorkom, J., 2005, \aap, 437, L19

\bibitem[Palunas \& Williams(2000)]{Palunas2000}
Palunas, P., \& Williams, T. B., \aj, 120, 2884

\bibitem[Poggianti et al.(1999)]{Poggianti1999}
Poggianti, B. M., et al. 1999, \apj, 518, 576

\bibitem[Possianti et al.(2001)]{Possiganti2001}
Poggianti, B. M., et al. 2001, \apj, 562, 689

\bibitem[Poggianti et al.(2004)]{Poggianti2004}
Poggianti, B. M., et al. 2004, \apj, 601, 197

\bibitem[Popesso et al.(2005)]{Popesso2005}
Popesso, P., Bohringer, H., Romaniello, M., \& Vogas, W. 2005, \aap, 433, 415

\bibitem[Popesso et al.(2006)]{Popesso2006}
Popesso, P., Biviano, A., Bohringer, H., \& Romaniello, M. 2006, \aap, 445, 29

\bibitem[Postman \& Geller(1984)]{Postman1984}
Postman, M., \& Geller, M. J. 1984, \apj, 281, 95

\bibitem[Postman et al.(2005)]{Postman2005}
Postman, M., et al. 2005, \apj, 623, 721

\bibitem[Quilis, Moore \& Bower(2000)]{Quilis2000}
Quilis, V., Moore, B., \& Bower, R. 2000, Science, 288, 1617

\bibitem[Roediger \& Hensler(2005)]{Roediger2005}
Roediger, E., \& Hensler, G. 2005, \aap, 433, 875

\bibitem[Roediger \& Br$\ddot{\rm u}$ggen(2007)]{Roediger2007}
Roediger, E., \& Br$\ddot{\rm u}$ggen, M. 2007, \mnras, 380, 1399

\bibitem[Sakai et al.(2002)]{Sakai2002}
Sakai, S., et al. 2002, \apj, 578, 842

\bibitem[Schlegel, Finkbeiner \& Davis(1998)]{Schlegel1998}
Schlegel, D. J., Finkbeiner, D. P., \& Davis, M. 1998, \apj, 500, 525

\bibitem[Schweizer(1978)]{Schweizer1978}
Schweizer, F. 1978, in Structure and Properties of Nearby Galaxies, p. 279

\bibitem[Schulz \& Struck(2001)]{Schulz2001}
Schulz, S., \& Struck, C. 2001, \mnras, 328, 185

\bibitem[Smith et al.(2005)]{Smith2005}
Smith, G. P., Treu, T., Ellis, R. S., Moran, S. M., \& Dressler, A. 2005, \apj, 620, 78

\bibitem[Struck \& Brown(2004)]{Struck2004}
Struck, C., \& Brown, J. R. 2004, in Duc, P.-A., Braine, J. \& Brinks, E., eds., Recycling 
Intergalactic and Interstellar Matter, IAU Symposium No. 217, p. 466

\bibitem[Sun, Donahue \& Voit(2007)]{Sun2007}
Sun, M., Donahue, M., \& Voit, M. 2007, \apj, 671, 170

\bibitem[Tran et al.(2003)]{Tran2003}
Tran, H. D., et al. 2003, \apj, 585, 750

\bibitem[Trentham, Sampson \& Banerji(2005)]{Trentham2005}
Trentham, N., Sampson, L., \& Banerji, M. 2005, \mnras, 357, 783

\bibitem[van der Wel et al.(2007)]{VanDerWel2007}
van der Wel, A., et al. 2007, \apj, 670, 206

\bibitem[Vollmer \& Huchtmeier(2003)]{Vollmer2003}
Vollmer, B., \& Huchtmeier, W. 2003, \aap, 406, 427

\bibitem[Vollmer et al.(2001)]{Vollmer2001}
Vollmer, B., Cayatte, V., Balkowski, C., \& Duschl, W. J. 2001, \apj, 561, 708

\bibitem[Vollmer et al.(2004)]{Vollmer2004}
Vollmer, B., Beck, R., Kenney, J. D. P., \& van Gorkom, J. H. 2004, \aj, 127, 3375

\bibitem[Yagi et al.(2007)]{Yagi2007}
Yagi, M., et al. 2007, \apj, 660, 1209

\bibitem[Yamanoi et al.(2007)]{Yamanoi2007}
Yamanoi, H., et al. 2007, \aj, 134, 56

\bibitem[Yoshida et al.(2002)]{Yoshida2002}
Yoshida, M., et al. 2002, \apj, 567, 118

\bibitem[Yoshida et al.(2004)]{Yoshida2004}
Yoshida, M., et al. 2004, \aj, 127, 90

\bibitem[Yoshino \& Ichikawa.(2008)]{Yoshino2008}
Yoshino, A., \& Ichikawa, T. 2008, \pasj, 60, 493

\end{thebibliography}
\end{document}